\documentclass[%
 reprint,
 amsmath,amssymb,
 aps,
 prl,
 superscriptaddress,
]{revtex4-2}

\usepackage[utf8]{inputenc}
\usepackage[T1]{fontenc}

\usepackage{amsthm}
\usepackage{txfonts} 
\usepackage{bm}
\usepackage{bbm}
\usepackage{mathrsfs}
\usepackage{tensor}

\usepackage[final]{graphicx}
\usepackage{dcolumn}
\usepackage{booktabs}
\usepackage{pifont}

\usepackage{color}

\definecolor{BLACK}{gray}{0}
\definecolor{WHITE}{gray}{1}
\definecolor{RED}{rgb}{1,0,0}
\definecolor{GREEN}{rgb}{0,1,0}
\definecolor{BLUE}{rgb}{0,0,1}
\definecolor{CYAN}{cmyk}{1,0,0,0}
\definecolor{MAGENTA}{cmyk}{0,1,0,0}
\definecolor{YELLOW}{cmyk}{0,0,1,0}

\providecommand{\U}[1]{\protect\rule{.1in}{.1in}}

\usepackage[
    colorlinks=true,
    citecolor=blue,
    linkcolor=blue,
    urlcolor=blue,
    anchorcolor=blue
]{hyperref}

\begin{document}

\title{Rotational Doppler Effect and Helimagnon Frequency Comb in Sliding Magnetic Spirals}
\author{Yang Liu}
\author{Jie Liu}
\author{Tianci Gong}
\author{Zhaozhuo Zeng}
\author{Xiaoping Gao}
\affiliation{School of Physics and State Key Laboratory of Electronic Thin Films and Integrated Devices, University of Electronic Science and Technology of China, Chengdu 610054, China}

\author{Kai Chang}
\email[Contact author: ]{kchang@zju.edu.cn}
\affiliation{Center for Quantum Matter, School of Physics, Zhejiang University, Hangzhou 310058, China}

\author{Peng Yan}
\email[Contact author: ]{yan@uestc.edu.cn}
\affiliation{School of Physics and State Key Laboratory of Electronic Thin Films and Integrated Devices, University of Electronic Science and Technology of China, Chengdu 610054, China}

\begin{abstract}
Collective spin motion can reshape magnon spectra by providing a dynamical magnetic background. Magnetic spirals offer a particularly clean setting, as they simultaneously host propagating helimagnons and a gapless Goldstone mode, i.e., the sliding of the spiral. Here, we show that once a helimagnon is excited and the spiral slides, the motion, which rotates the local spin frame, produces a thresholdless rotational Doppler shift of the helimagnon spectrum, with even winding for the linear mode and odd winding for its second harmonic. Stronger excitation generates one finite-momentum phason. Mixing of that phason with the helimagnon produces a helimagnon frequency comb only inside a finite driving window, spaced by the phason frequency fixed by the spiral length rather than by the sliding rate. These two scales are separable in a microwave experiment: the Doppler offset tracks the sliding velocity, whereas the comb spacing scales inversely with the length of the spiral.
\end{abstract}

\maketitle



\textit{Introduction}---Magnetic spirals are periodically twisted spin textures, stabilized by the competition between Heisenberg exchange and Dzyaloshinskii--Moriya interaction \cite{Dzyaloshinskii1958, Moriya1960, BakJensen1980, Uchida2006, PhysRevLett.107.017205, PhysRevB.100.060404, PhysRevB.106.205110, Baral2025}. The twist folds the spin-wave spectrum into helimagnon bands, with chiral resonances and a nonreciprocal dispersion \cite{Belitz2006, PhysRevB.79.220405, Koralek2012, Schwarze2015, Kugler2015, Weiler2017, Iguchi2017, Ogawa2021}. It also breaks continuous translation along the spiral axis. The resulting gapless Goldstone mode is the sliding of the spiral \cite{RevModPhys.85.219, Kirkpatrick2005, Ho2010, PhysRevB.87.174409, Hongo2021, tqqw-qh7l}: a displacement along the axis is locked to a rotation of the local spin frame \cite{Kishine2012, DelSer2021}. A time-dependent magnetic field drives this sliding. The velocity is second order in the drive, is resonantly enhanced when a helimagnon is excited, and reverses with the driven branch \cite{Kishine2012, Goncalves2017, DelSer2021}. The same branch at finite wave vector describes phasons, the phase fluctuations of the spiral. Helimagnon resonances and these phase fluctuations have been resolved spectroscopically, and the phason dispersion is chiral \cite{Goncalves2017, Shimamoto2022, PhysRevB.111.024411, Wu2025}. A steady sliding advances the spiral phase and rotates the frame in which a helimagnon is defined, so the sliding is a time-dependent magnetic background for the helimagnon spectrum. What has not been established is how that sliding shifts the helimagnon frequencies, and how a finite-momentum phason, rather than the sliding itself, sets a separate frequency scale in the spectrum.

\begin{figure}[b]
    \centering
    \includegraphics[width=0.48\textwidth]{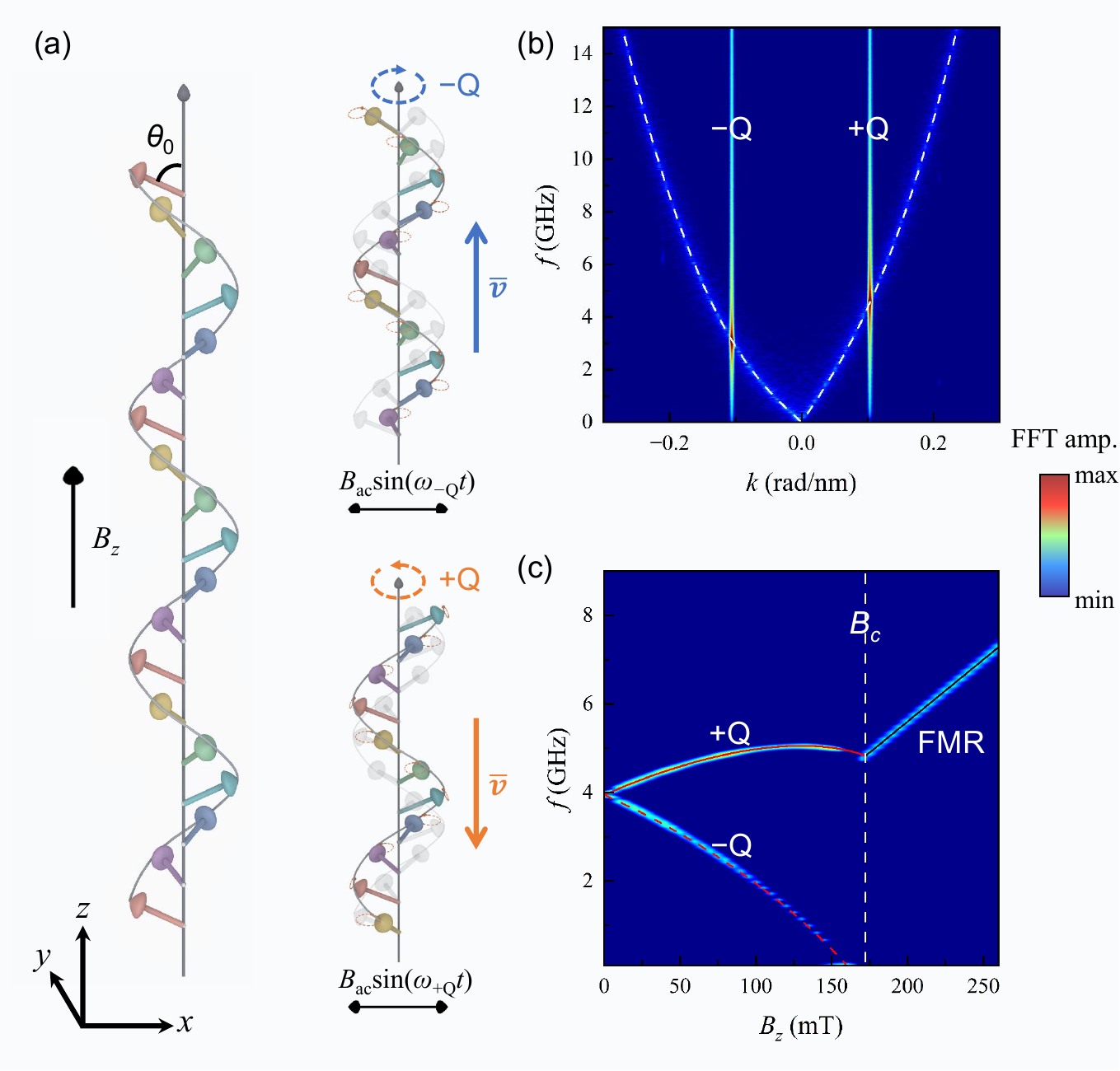}
    \caption{Selective excitation of $\pm Q$ helimagnons and spiral sliding in a conical spin spiral.
    (a) Schematic of a conical spiral under a longitudinal static field $B_z$. A transverse microwave field of amplitude $B_{\rm ac}$ and frequency $\omega_{\rm ac}=2\pi f_{\rm ac}$ is tuned to $\omega_{-Q}$ or to $\omega_{+Q}$, resonantly exciting the elliptically polarized helimagnon modes at $k=-Q$ or $k=+Q$, respectively, with collective spiral sliding along opposite directions owing to the second-order rectification.
    (b) Simulated magnon spectrum at $B_z=50~\mathrm{mT}$, with the theoretical dispersions superimposed (dashed curves).
    (c) Field-dependence of the $\pm Q$ helimagnon frequencies. The vertical dashed line denotes the critical field $B_c$, above which the conical spiral transforms into a uniform ferromagnetic state and the ferromagnetic resonance (FMR) mode emerges. Curves are analytical results.
}
    \label{fig1}
\end{figure}

In this Letter, we show how that sliding is written onto the helimagnon spectrum. Because the sliding rotates the local spin frame, it produces a rotational Doppler shift of the helimagnon frequencies. The shift has no threshold of its own. The winding is even for the linear mode and odd for the second harmonic. At stronger excitation the driven helimagnon generates one finite-momentum phason. Mixing of that phason with the helimagnon produces a helimagnon frequency comb. Unlike the Doppler shift, the comb exists only inside a finite driving window, and its spacing is the phason frequency fixed by the length of the spiral. Both signatures lie in the gigahertz range and can be read by microwave spectroscopy on spirals of controlled length.

\textit{Helimagnon excitation and spiral sliding}---We consider a chiral magnet described with the following Hamiltonian
    \begin{align}
    \mathcal{H}
    =
    A(\partial_z\mathbf{m})^2
    +
    D\mathbf{m}\cdot(\nabla\times\mathbf{m})
    -
    M_s\mathbf{B}\cdot\mathbf{m},
    \label{eq:model}
    \end{align}
where $\mathbf{m}(z,t)$ is the unit magnetization, $A$ is the Heisenberg exchange coefficient, $D$ is the bulk Dzyaloshinskii-Moriya interaction (DMI) constant, and $M_s$ is the saturation magnetization. A static magnetic field $B_z\hat{\mathbf z}$ applied along the helical axis stabilizes a conical spiral \cite{PhysRevB.68.064406, PhysRevB.92.125110, PhysRevB.99.064414, PhysRevB.106.064417},
    \begin{align}
    \mathbf{m}_0(z,t)
    =
    \big[
    \sin\theta_0\cos\Phi(z,t),
    \sin\theta_0\sin\Phi(z,t),
    \cos\theta_0
    \big],
    \label{eq:conical_state}
    \end{align}
where $\Phi(z,t)$ denotes the spiral phase and $\theta_0$ is the cone angle, as shown in Fig.~\ref{fig1}(a). In the rigid-spiral approximation, the collective translation of the spiral is described by $\Phi(z,t)=Q[z-Z(t)]$, where the pitch $Q$ is fixed by the boundary condition and $Z(t)$ is the displacement along the helical axis ($z$ axis) \cite{PhysRevLett.30.230, PhysRevB.78.064425, PhysRevB.93.054403, PhysRevB.101.184425}. Minimizing the static energy at that fixed $Q$ gives $2(DQ-AQ^2)\cos\theta_0=M_sB_z$ (see Sec. I of the Supplemental Material \cite{SM}). The conical phase is stable below the critical field $B_c=2(DQ-AQ^2)/M_s$ and continuously reaches the field-polarized state, i.e. $\theta_0\rightarrow0$, as $B_z$ increases.

To determine the helimagnon spectrum, we expand the magnetization in the local frame of the conical spiral as $\mathbf{m}=\hat{\mathbf e}_r+\delta\theta \hat{\mathbf e}_{\theta}+\delta\phi \hat{\mathbf e}_{\phi}$, where $\delta\theta$ and $\delta\phi$ denote the two transverse components around the equilibrium spiral \cite{PhysRevLett.107.177207, nlgc-rh7s}. For helimagnons with frequency $\omega$ and wavevector $k$, we obtain
    \begin{align}
    \delta\theta(z,t)
    &=
    \sum_{\sigma=\pm1}
    \operatorname{Re}
    \left[
    \mathcal{U}_\sigma
    e^{ikz-i\omega t}
    \right],
    \notag\\
    \delta\phi(z,t)
    &=
    \sum_{\sigma=\pm1}
    \operatorname{Re}
    \left[
    \mathcal{V}_\sigma
    e^{ikz-i\omega t}
    \right],
    \label{eq:helimagnon_fluctuations}
    \end{align}
where $\mathcal{U}_\sigma$ and $\mathcal{V}_\sigma$ are the corresponding complex transverse amplitudes that encode both the magnitude and phase, and $\sigma=\pm1$ denotes the $\pm Q$ helimagnon branch, respectively. Linearizing the Landau-Lifshitz-Gilbert equation yields the helimagnon dispersion (see Sec.~II \cite{SM} for detailed derivations, where Gilbert damping was omitted)
    \begin{align}
    \omega(k)
    =
    \frac{2}{s}
    \Bigg\{
    &k(D-2AQ)\cos\theta_0
    \notag\\
    &+
    \sqrt{
    Ak^2
    \left[
    Ak^2+(DQ-AQ^2)\sin^2\theta_0
    \right]
    }
    \Bigg\},
    \label{eq:helimagnon_dispersion}
    \end{align}where $s=M_s/\gamma$ is the spin density with $\gamma$ being the gyromagnetic ratio. The term odd in $k$ produces a nonreciprocal spectrum for counterpropagating helimagnons. Using material parameters of Cu$_2$OSeO$_3$ \cite{Chen2026, Wu2025}, $A=3.6\times10^{-13}~\mathrm{J/m}$, $D=1.2\times10^{-4}~\mathrm{J/m^2}$, and $M_s=1.0\times10^{5}~\mathrm{A/m}$, we calculate the helimagnon spectrum and perform micromagnetic simulations using the MuMax3 package \cite{mumax}. The analytical dispersion obtained from Eq.~\eqref{eq:helimagnon_dispersion} agrees well with the simulation results [Fig.~\ref{fig1}(b)], where particularly strong spectral weight appears around the $k=\pm Q$ helimagnon modes driven by the uniform sinc-function field. Their resonance frequencies as functions of $B_z$ are shown in Fig.~\ref{fig1}(c). The applied field lifts the degeneracy between the $+Q$ and $-Q$ modes. As $B_z$ approaches the critical field $B_c$, the conical state evolves into the field-polarized ferromagnetic state, and the helimagnon excitation recovers the conventional FMR mode above $B_c$, with the frequency given by $f_{\rm FMR}=\gamma B_z/2\pi$.
    
    \begin{figure}[t]
        \centering
        \includegraphics[width=0.48\textwidth]{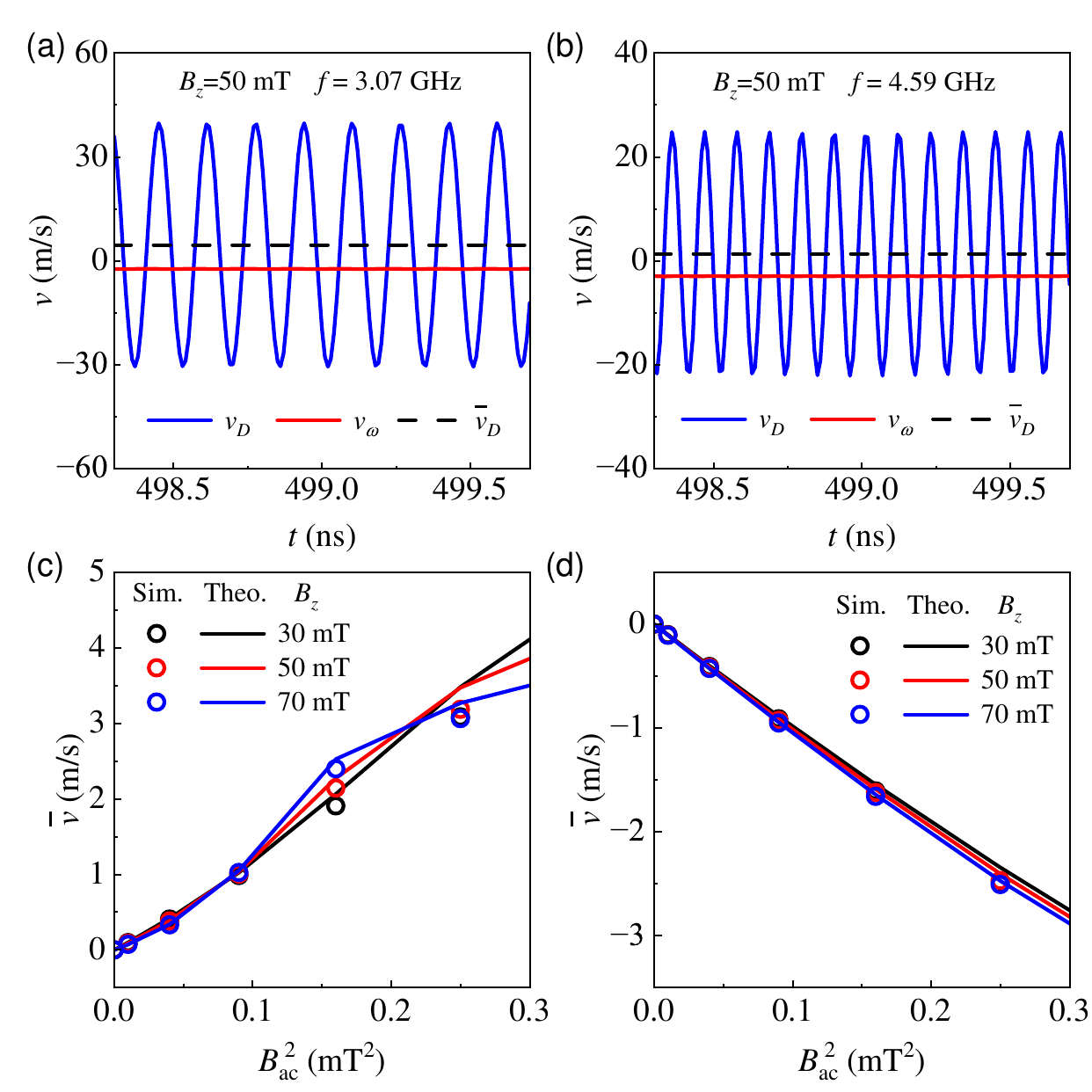}
        \caption{Microwave-driven spiral sliding. (a), (b) Time-dependent $v_D$ (blue curves) and $v_{\omega}$ (red curves) for two representative helimagnon excitations producing opposite sliding directions. The black dashed line denotes the time-averaged contribution $\overline{v}_D$. (c), (d) Time-averaged spiral velocity $\overline{v}$ as a function of the squared microwave-field amplitude $B_{\rm ac}^{2}$ for $B_z=30$, $50$, and $70$~mT, for the two excitation frequencies in (a) and (b), respectively. Open circles denote micromagnetic simulations ($\alpha = 0.01$), and solid curves represent the analytical results obtained from Eq.~\eqref{eq:sliding_velocity}.}
        \label{fig2}
    \end{figure}

The nonreciprocal helimagnon dispersion described by Eq.~\eqref{eq:helimagnon_dispersion} enables selective excitation of the two helimagnon branches by merely tuning the microwave frequency. Specifically, excitation at $\omega_{+Q}$ and $\omega_{-Q}$ selectively addresses the $+Q$ and $-Q$ helimagnon mode, respectively, resulting in spiral sliding in opposite directions, as illustrated in Fig.~\ref{fig1}(a). We derive the analytical formula of the sliding velocity
    \begin{align}
    v_\sigma = v_{D,\sigma} + v_{\omega,\sigma}.
    \label{eq:sliding_velocity}
    \end{align}
Here, $v_{D,\sigma}=\frac{D}{\alpha s\cos\theta_0} \operatorname{Re} \left(\mathcal{U}_{\sigma}\mathcal{V}_{\sigma}^{*}\right)$ originates from the in-phase product of the two transverse components, and $v_{\omega,\sigma} = \frac{\omega}{2Q\cos\theta_0}\operatorname{Im}\left( \mathcal{U}_{\sigma}\mathcal{V}_{\sigma}^{*} \right)$ from their quadrature. Writing $\mathcal{U}_{\sigma} = |\mathcal{U}_{\sigma}|e^{i\delta_{\theta,\sigma}}$ and $\mathcal{V}_{\sigma} = |\mathcal{V}_{\sigma}|e^{i\delta_{\phi,\sigma}}$, and defining the relative phase as $\Delta_{\sigma} = \delta_{\theta,\sigma} - \delta_{\phi,\sigma}$, their time-average can be expressed as (see Secs. III and IV \cite{SM})
    \begin{subequations}\label{eq6}
    \begin{equation}\label{eq6a}
    \overline{v}_{D,\sigma}
    =
    |\mathcal{U}_{\sigma}\mathcal{V}_{\sigma}|
    \frac{D}{\alpha s\cos\theta_0}
    \cos\Delta_{\sigma},
    \end{equation}
    \vspace{-2em} 
    \begin{equation}\label{eq6b}
    \overline{v}_{\omega,\sigma}
    =
    |\mathcal{U}_{\sigma}\mathcal{V}_{\sigma}|
    \frac{\omega}{2Q\cos\theta_0}
    \sin\Delta_{\sigma}.
    \end{equation}
    \end{subequations}
Equations~\eqref{eq6} show that the sliding direction is determined not only by the helimagnon intensity, but also by its polarization and, in particular, by the relative phase between the two transverse components. The two contributions to the sliding dynamics are illustrated in Figs.~\ref{fig2}(a) and \ref{fig2}(b): $v_D$ exhibits a pronounced oscillation under the microwave drive, while its time average $\overline{v}_D$ remains finite. In contrast, $v_{\omega}$ keeps a constant value without oscillation. For the $-Q$ mode in Fig.~\ref{fig2}(a), the positive $\overline{v}_D$ dominates over the negative $\overline{v}_{\omega}$, resulting in a positive net sliding velocity. For the $+Q$ mode in Fig.~\ref{fig2}(b), the negative $v_{\omega}$ becomes dominant instead, leading to a negative sliding. The reversal of the sliding direction is closely related to the different polarizations and relative phases of helimagnons. Detailed analyses show that the relative phase $\Delta_{\sigma}$ is sensitive to damping, but not to the microwave amplitude (see Sec.~IV \cite{SM}).

Another consequence dictated by Eqs.~\eqref{eq6} is the quadratic dependence of the sliding velocity on the microwave amplitude. In the linear response, both $\mathcal{U}_{\sigma}$ and $\mathcal{V}_{\sigma}$ are proportional to $B_{\rm ac}$, whereas their amplitude ratio and relative phase are not (see Sec.~III \cite{SM}). The time-averaged velocity is therefore quadratic $\overline{v}\propto \mathcal{U}_{\sigma}\mathcal{V}_{\sigma}^{*}\propto B_{\rm ac}^{2}$. Figures~\ref{fig2}(c) and \ref{fig2}(d) confirm this for both branches and for three values of $B_z$. Changing $B_z$ changes the cone angle and the polarization, and hence the magnitude of $\overline{v}$. The quadratic law and the reversal with the driven branch are the Archimedean-screw response \cite{DelSer2021}. The expressions above identify its two channels for each helimagnon branch.

    \begin{figure}[t]
        \centering
        \includegraphics[width=0.5\textwidth]{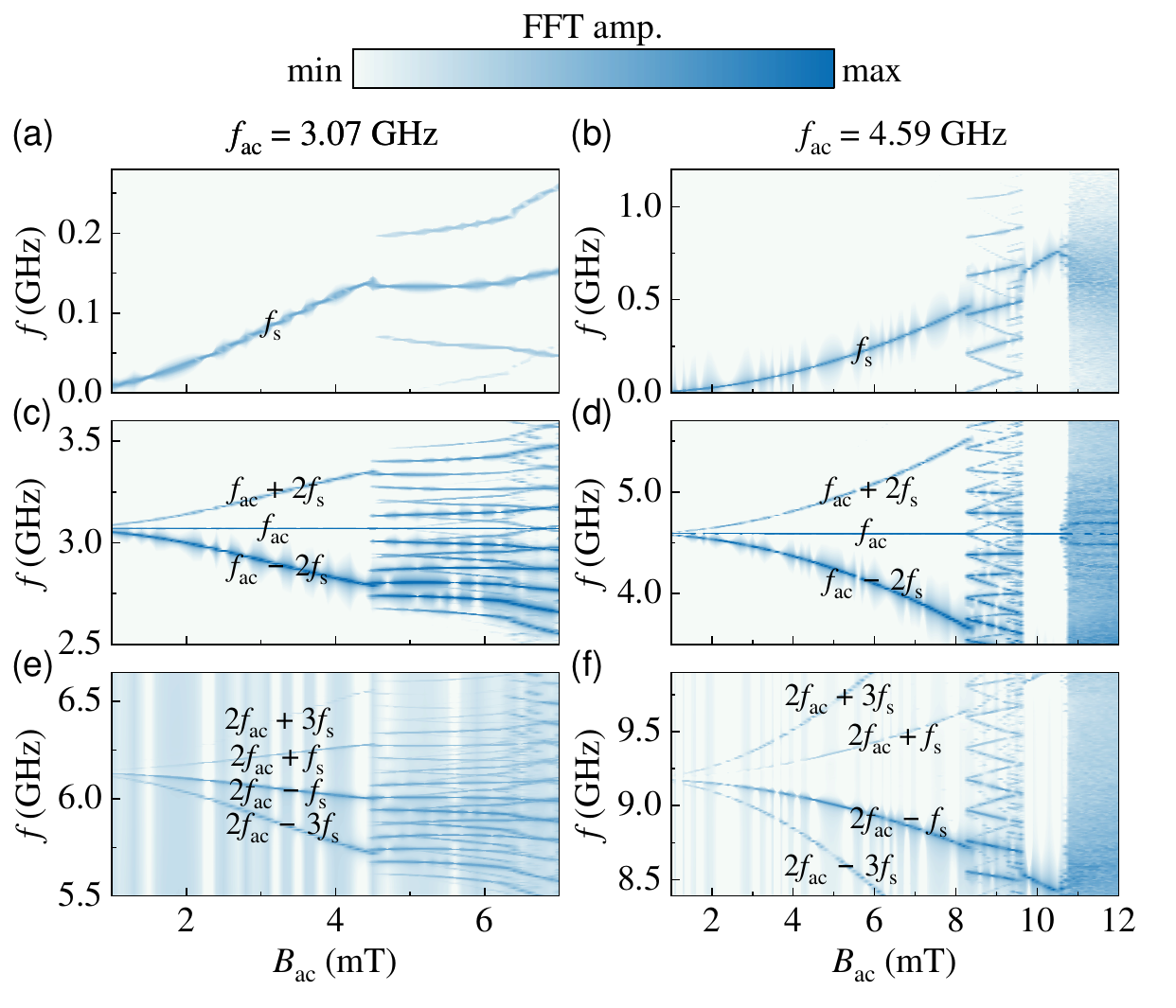}
        \caption{
        Microwave-induced spiral sliding and the resulting rotational Doppler spectra at $B_z=50~\mathrm{mT}$.
        (a), (b) Low-frequency spectra for $f_{\rm ac}=3.07$ and $4.59~\mathrm{GHz}$, respectively. The branch labeled $f_{\rm s}$ corresponds to the spiral sliding frequency. (c), (d) High-frequency spectra around the fundamental excitation frequency, showing the branches $f_{\rm ac}$ and $f_{\rm ac}\pm2f_{\rm s}$.
        (e), (f) Spectra around the second harmonic, i.e., 2$f_{\text{ac}}$, where the characteristic branches occur at $2f_{\rm ac}\pm f_{\rm s}$ and $2f_{\rm ac}\pm3f_{\rm s}$. The color scale represents the FFT amplitude of $m_y$ component for a damping constant of $\alpha=0.05$.}
        \label{fig3}
    \end{figure}

\textit{Rotational Doppler effect and helimagnon frequency comb}---Next, we show that the collective spiral sliding couples directly to the phase winding of helimagnon excitations, providing a natural route to a rotational Doppler shift \cite{Garetz1981, Courtial1998, PhysRevA.82.052506, PhysRevLett.134.137001, Lendinez2010, Chudo2021, Wang2022}. For a steady sliding, we have
    \begin{equation}
    \Phi(z,t)
    =
    Q[z-Z(t)]
    =
    Qz-\Omega_{\rm s}t,
   \ \ 
    \text{with}\ \ \Omega_{\rm s}
    =
    Q\overline{v},
    \label{eq:RDE_phase}
    \end{equation}
where $\Omega_{\rm s}$ is the rotation frequency of the spiral phase $\Phi$. As shown in Figs.~\ref{fig3}(a) and \ref{fig3}(b), a pronounced low-frequency branch shifts continuously to higher frequencies with increasing $B_{\rm ac}$. Its frequency closely follows $f_{\rm s}=\Omega_{\rm s}/2\pi$, identifying this branch as the sliding-induced spiral phase rotation. We then adopt the winding number $n$ to characterize the spiral phase, namely, $e^{in\Phi}$. For the sliding spiral in Eq.~\eqref{eq:RDE_phase}, such a component evolves as
    \begin{equation}
    e^{-i\omega_{\rm ac} t}e^{in\Phi(z,t)}
    =
    e^{inQz}
    e^{-i(\omega_{\rm ac}+n\Omega_{\rm s})t},
    \label{eq:winding_shift}
    \end{equation}
which shows that a helimagnon with spiral-phase winding $n$ acquires a rotational Doppler shift $n\Omega_{\rm s}$. Below, we illustrate this phenomenon and explain the origin of the winding number.

The driven helimagnon on branch $\sigma$ already carries an intrinsic spiral-phase winding $\sigma$, while the transformation from the local frame to laboratory frame generates an additional geometrical winding $+1$, $-1$, and $0$ to $m_+=m_x+ i m_y$, $m_-=m_x- i m_y$, and $m_z$, respectively. The transverse windings are therefore the even numbers $n=\sigma\pm 1$, and $m_z$ carries the odd winding $n=\sigma$. This can be summarized by the following winding matrix (see Sec.~V \cite{SM})
    \begin{equation}
    \hat{\mathcal N}_{\sigma}
    =
    \begin{pmatrix}
    \sigma+1 & 0 & 0\\
    0 & \sigma-1 & 0\\
    0 & 0 & \sigma
    \end{pmatrix}.
    \label{eq:winding_matrix}
    \end{equation}
Then, we obtain the frequency matrix
    \begin{equation}
    \hat{\Omega}_{\sigma}^{\rm lab}
    =
    \begin{pmatrix}
    \omega_{\rm ac}+(\sigma+1)\Omega_{\rm s} & 0 & 0\\
    0 & \omega_{\rm ac}+(\sigma-1)\Omega_{\rm s} & 0\\
    0 & 0 & \omega_{\rm ac}+\sigma\Omega_{\rm s}
    \end{pmatrix}.
    \label{eq:frequency_matrix_explicit}
    \end{equation}
In laboratory frame, the helimagnon dynamics can be written compactly as
    \begin{equation}
    \mathbf m_{\sigma}
    =
    e^{-i\hat{\Omega}_{\sigma}^{\rm lab}t}
    \mathbf A_{\sigma},
    \label{eq:lab_dynamics_matrix}
    \end{equation}
 where $\mathbf A_{\sigma} = \left( \mathcal U_{\sigma}\cos\theta_0+i\mathcal V_{\sigma}, ~ \mathcal U_{\sigma}\cos\theta_0-i\mathcal V_{\sigma}, ~ -\mathcal U_{\sigma}\sin\theta_0 \right)^{\text T}$ is the helimagnon amplitude vector for branch $\sigma$. The sliding-induced phase rotation therefore produces component-dependent rotational Doppler shifts determined by the total spiral-phase winding. For linear helimagnons excited by microwave with frequency $f_{\rm ac}$, their spectra contain three branches, $f_{\rm ac}-2f_{\rm s}$, $f_{\rm ac}$, and $f_{\rm ac}+2f_{\rm s}$. These branches are resolved in Figs.~\ref{fig3}(c) and \ref{fig3}(d). The shift of the two sidebands (i.e., $f_{\rm ac}\pm 2f_{\text s}$) from the central branch increases continuously with $B_{\rm ac}$ and follows $2f_{\rm s}$, providing numerical evidence of the rotational Doppler effect.

Then, we show that the winding-number mechanism applies to helimagnon higher-harmonics too. An $r$th-order nonlinear contribution involves a product of $r$ helimagnon mode amplitudes, with the $\ell$th factor belonging to the branch $\sigma_\ell=\pm1$ ($\ell=1,\ldots,r$). The total intrinsic winding number is then $S=\sum_{\ell=1}^{r}\sigma_\ell=r-2j$, where $j=0,1,\ldots,r$ labels the number of components belonging to the $\sigma_\ell=-1$ branch. Including the geometrical winding due to the rotation of reference frame gives (see Sec.~V \cite{SM})
    \begin{equation}
    \hat{\mathcal N}_{S}
    =
    \begin{pmatrix}
    S+1 & 0 & 0\\
    0 & S-1 & 0\\
    0 & 0 & S
    \end{pmatrix}.
    \label{eq:nonlinear_winding_matrix}
    \end{equation}
The frequency matrix for the $r$th-order process is then $\hat{\Omega}_{r,S}^{\rm lab} = r\omega_{\rm ac}\mathbf I_3 + \Omega_{\rm s}\hat{\mathcal N}_{S}$, where $\mathbf I_3$ is the $3\times3$ identity matrix. Consequently, both $x-$ and $y-$components of the helimagnon oscillate with the following frequency
    \begin{equation}
    f_{r, j}^{\rm lab,\pm}
    =
    r f_{\rm ac}
    +
    (r-2j\pm1)f_{\rm s}.
    \label{eq:nonlinear_RDE}
    \end{equation}
The second-harmonic spectra in Figs.~\ref{fig3}(e) and \ref{fig3}(f) clearly illustrate this nonlinear winding structure. For $r=2$, Eq.~\eqref{eq:nonlinear_RDE} predicts four branches $2f_{\rm ac}-3f_{\rm s}$, $2f_{\rm ac}-f_{\rm s}$, $2f_{\rm ac}+f_{\rm s}$, and $2f_{\rm ac}+3f_{\rm s}$, fully consistent with numerical simulations. Notably, the fundamental response (i.e., $r=1$) is associated with the even winding numbers $n=0,\pm2$, whereas the second-harmonic response merely carries the odd winding numbers $n=\pm1,\pm3$. This even--odd contrast demonstrates that the microwave-induced spiral sliding imprints a hierarchy of rotational Doppler sidebands onto the helimagnon spectrum.

\begin{figure}[t]
        \centering
        \includegraphics[width=0.5\textwidth]{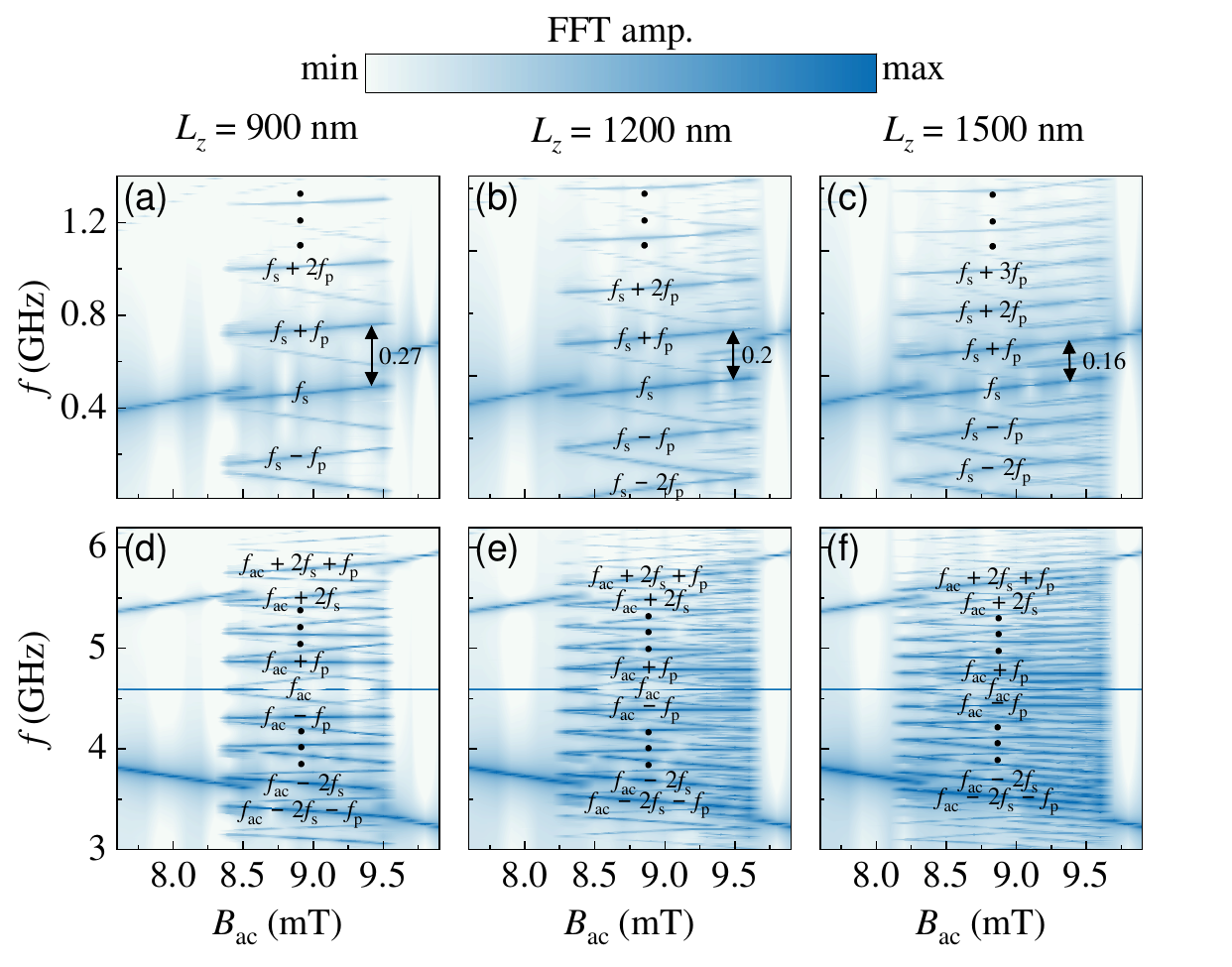}
        \caption{ Phason-induced helimagnon frequency combs for spiral length $L_z=900$, $1200$, and $1500~\mathrm{nm}$ at $B_z=50~\mathrm{mT}$ and $f_{\rm ac}=4.59~\mathrm{GHz}$. (a)--(c) Low-frequency spectra around the spiral-sliding branch. The characteristic frequency separation decreases from $0.28$ to $0.20$ and $0.16~\mathrm{GHz}$ with increasing $L_z$, consistent with the scaling $f_{\rm p}=f_{{\rm p},1}^{+1}\propto 1/L_z$.
        (d)--(f) High-frequency spectra around the fundamental helimagnon response, showing phason-induced fine structures around $f_{\rm ac}$ and the rotational-Doppler-shifted branches $f_{\rm ac}\pm2f_{\rm s}$. The color scale represents the FFT amplitude of $m_y$ component for a damping constant of $\alpha=0.05$.}
        \label{fig4}
    \end{figure}

At larger microwave amplitude, a finer set of lines appears, as shown in Figs. \ref{fig3}(a), \ref{fig3}(c), \ref{fig3}(e) for $B_{\rm ac}>4.2$ mT and Figs. \ref{fig3}(b), \ref{fig3}(d), \ref{fig3}(f) for $B_{\rm ac}>8$ mT, forming a distinct helimagnon frequency comb \cite{MFC1, Wang2022}. We show that these spectral structures are not produced by the uniform sliding, but by the quantized phase fluctuations \cite{Togawa2015, Kishine2019}, which modulate both the spiral-sliding mode and the rotational-Doppler-shifted helimagnon mode. To this end, we express the instantaneous spiral phase as
    \begin{equation}
    \Phi(z,t)
    =
    Q[z-Z(t)]+\delta\Phi(z,t),
    \label{eq:total_phase_phason}
    \end{equation}
where $\delta\Phi$ denotes the phase fluctuation. In the long-wavelength limit, the phason dispersion takes the form (see Sec. VI \cite{SM})
    \begin{equation}
    \omega_{\rm p}(q)
    =
    \frac{2\sin\theta_0\sqrt{A(DQ-AQ^2)}}{s}|q|
    +
    \frac{2(D-2AQ)\cos\theta_0}{s}q ,
    \label{eq:phason_dispersion}
    \end{equation}
where $q$ is the phason wavevector. The second term is odd in $q$, generating the nonreciprocity, i.e., $\omega_{\rm p}(+q)\neq\omega_{\rm p}(-q)$. For a spiral magnet of a finite length $L_z$, the allowed phason wave vectors are quantized as $q_n=2\pi n/L_z$ with $n=1,2,3,\ldots$ by imposing the periodic boundary condition. We then denote the chirality-dependent phason dispersion as $\omega_{{\rm p},n}^{\rho}=\omega_{\rm p}(\rho q_n)$ with $\rho=\pm1$. The linear phason dispersion \eqref{eq:phason_dispersion} gives $f_{{\rm p},1}^{\rho}= \omega_{{\rm p},1}^{\rho}/2\pi\propto 1/L_z$. This size dependence can be clearly resolved in the low-frequency spectra of Figs.~\ref{fig4}(a)-\ref{fig4}(c). For $L_z=900$, $1200$, and $1500~\mathrm{nm}$, the characteristic frequency separations $\Delta f$ are $0.28$, $0.20$, and $0.16~\mathrm{GHz}$, respectively. The product $L_z\Delta f$ is approximately constant, and the spacing follows the $\rho=+1$ branch selected by that drive. 

The comb occupies only a finite interval of $B_{\rm ac}$. For $L_z=1200~\mathrm{nm}$ the phason amplitude is still small at $8.00~\mathrm{mT}$ and has grown and saturated by $8.50~\mathrm{mT}$ (see Fig. S3 in Sec.~VII \cite{SM}). The rotational Doppler shift has no such threshold. The interval is fixed by the decay of the driven helimagnon into one phason and one lower sideband. To gain analytical understanding, we introduce the instability discriminant of the phason (Sec.~IX \cite{SM})
    \begin{equation}
    F(B_{\rm ac})=g^{2}-\kappa^{2}-[\epsilon(B_{\rm ac})]^{2}/4,
    \end{equation}
where $g=g_1B_{\rm ac}$ is the rate at which the driven helimagnon creates that phason and sideband, with $g_1$ set by the three-magnon vertex and by the linear response of the driven mode, $\kappa$ is the common amplitude decay of the phason and the sideband, and $\epsilon$ is the detuning from $\omega_{\rm p}+\omega_l=\omega_{\rm ac}$, with $\omega_l$ the frequency of the sideband emitted together with the phason. Through the simulated window, $\epsilon(B_{\rm ac})=\epsilon_0+\chi_2B_{\rm ac}^{2}+\chi_4B_{\rm ac}^{4},$
where $\epsilon_0$ is the detuning at zero drive, $\chi_2<0$, and $\chi_4>0$. When $F>0$ the phason amplitude grows, and when $F<0$ it decays. The negative $\chi_2$ reduces $\epsilon$ from $\epsilon_0$, but alone it makes $|\epsilon|$ grow before $g$ exceeds $\kappa$, so $F$ never becomes positive. The positive $\chi_4$ cancels most of that shift inside the window and dominates beyond it. Thus $F=0$ has two positive roots. They are the lower and upper critical fields. Here $g_1$ and $\epsilon_0$ depend on $L_z$, while $\kappa$, $\chi_2$, and $\chi_4$ do not. Their values, and the comparison with the simulated amplitude, are given in Sec. IX and Fig. S4 of the Supplemental Material \cite{SM}.

The emerging sidebands are then given by
    \begin{equation}
    \begin{aligned}
    f_{{s},n}^{\rho}
    &=f_{\rm s}+\rho f_{{\rm p},n}^{\rho}, \\
    f_{{h},\nu,n}^{\rho}
    &=f_{\rm ac}+2\nu f_{\rm s}+\rho f_{{\rm p},n}^{\rho},
    \end{aligned}
    \label{eq:phason_RDE_comb}
    \end{equation}
where $ f_{{\rm p},n}^{\rho}=\omega_{{\rm p},n}^{\rho}/2\pi$ and the subscripts $s$ and $h$ denote the phason sidebands around the sliding-frequency carrier $f_{\rm s}$ and the helimagnon rotational-Doppler carriers [Eq. \eqref{eq:nonlinear_RDE} with $r=1$], respectively. Here, $\nu=0,~\pm1$ labels the three helimagnon carriers. This structure is manifested in Fig.~\ref{fig4}: Figs.~\ref{fig4}(a)-\ref{fig4}(c) show the phason sidebands around the spiral-sliding branch $f_\text{s}$, while Figs.~\ref{fig4}(d)-\ref{fig4}(f) reveal the corresponding fine structures around the helimagnon carriers $f_{\rm ac}$ and $f_{\rm ac}\pm2f_{\rm s}$. It is noted that the phason-induced fine structure around the second-harmonic of helimagnon is too weak to be resolved, possibly due to the large damping adopted in our simulations.

\textit{Conclusion}---The dc sliding is the Archimedean-screw response already expected for a driven spiral, and its magnitude is fixed at second order by the coupling to the Goldstone mode. What the helimagnon spectrum adds is a rotational Doppler effect, with even winding in the linear spectrum and odd winding in its higher harmonics. Under stronger drive, mixing with one finite-momentum phason produces a helimagnon frequency comb whose spacing is inversely proportional to the spiral length and which appears only inside a finite driving window. Both frequencies lie in the gigahertz range and can be separated by microwave experiments on a conical spiral such as Cu$_{2}$OSeO$_{3}$, or on a monoaxial chiral magnet such as CrNb$_{3}$S$_{6}$. Our findings pave the way to controlling a magnon spectrum by the velocity and the length of a sliding spiral, independently of the frequency of the drive.

\textit{Acknowledgments}---This work was funded by the National Key R$\&$D Program of China (No. 2025YFA1411302 and No. 2022YFA1402802), the National Natural Science Foundation of China (NSFC) (Nos. 12625405, 12374103, 12434003), and Sichuan Science and Technology Program (No. 2025NSFJQ0045). J. L. acknowledges the financial support from NSFC (Grant No. 12604157). K. C. was supported by NSFC (Grant No. 12488101). Y. L. acknowledges the financial support from the China Postdoctoral Science Foundation (Grant No. 2025M773378). Z.Z. acknowledges the financial support from NSFC (Grant No. 125B2070).

\textit{Data availability}---The data that support the findings of this article are publicly available \cite{DA}.

\appendix


\end{document}